\documentclass[12pt]{article}

\input epsf

\title{ $K^{-}p$ and $\pi^{-}p$ charge-exchange processes \\
and $\eta - \eta'$ mixing}
\author{M. L. Nekrasov \\
{\small\it Institute for High Energy Physics, 142281 Protvino,
Russia}}

\date{}
\begin{document}
\maketitle

\begin{abstract}
The processes of charge-exchanges of $\pi^{-}$ and $K^{-}$ on protons at
high energies and zero transfers are described on the basis of the
parton model. It is shown that at the  mentioned kinematic conditions
the scattering occurs on the valence quarks only, while scattering on
sea quarks leads to disruption of final state. Thereby a substantiation
is given of the valence approximation for the specified reaction. On the
basis of GAMS-4$\pi$ data it is shown that the description of $\eta -
\eta'$ mixing does not require an introduction of additional states that
do not fit into the scheme of basis states $|\eta_N\!\!>$ and
$|\eta_S\!\!>$. The resulting value of the mixing angle is
$\phi_p=(36.1\pm 0.9)^0$.
\end{abstract}

\section{Introduction}\label{sec1}

The problem of $\eta-\eta'$ mixing is the oldest one existing almost
from the time that SU(3)-flavor symmetry was proposed. However, until
now this problem has no satisfactory solution \cite{PDG}. The reason
may lie in the model dependence of theoretical description of
particular processes \cite{Feldmann}, or  in an unknown global
systematic error admitted at the data analysis. Anyway the problem
manifests itself as instability in determining the mixing angle
\cite{BES}. Probable reason is sometimes associated with an admixture
of higher-mass states, primarily of a glueboll. However, in reality
this does not solve the problem, since inclusion of such states also
leads to ambiguous situation, see e.g. \cite{Feldmann} and recent
works \cite{Bele,ANTIKLOE,KLOE}.

In this connection it is of great interest an appearance of any
new reliable data to include them in the analysis. Recently the data
with high statistics were obtained on charge-exchange reactions
$\pi^{-}\,p \to \eta(\eta')\,n$ and $K^{-}\,p \to \eta(\eta')\Lambda$
at $32.5$ GeV/c. \cite{GAMS-4p}. Previously similar data were available
at different energies, but with relatively low statistics and only in
the case of the former reaction \cite{NICE,Stanton,ACCMOR}. The data
with $K^-$ beams were obtained for the first time, and this gave
additional advantages for studying the $\eta -\eta'$ mixing.
Unfortunately, the theoretical part of analysis of \cite{GAMS-4p}, in
our opinion, includes inaccuracies. The purpose of this Letter is to
correct detected inaccuracies, and to propose more complete development
of the model with the aid of which the mixing of states in the context
of given reactions can be investigated.

\section{Substantiation of the model of investigation}\label{sec2}

As a rule, the analysis of mixing of states is carried out in the
framework of the na\"{\i}ve quark model. The essence of this model
consists in an idea that observable particles are made up of minimal
allowed by quantum numbers set of constituent quarks and antiquarks. In
particular, isosinglet mesons are made up of neutral on flavors pairs of
constituent quarks and antiquarks. So, the $\eta$ and $\eta'$ are
considered as linear combinations of pairs of nonstrange and strange
quarks and antiquarks,
\begin{eqnarray}\label{text1}
\left( \begin{array}{c}
\eta  \\
\eta' \\
         \end{array} \right)
=
\left( \begin{array}{cc}
\cos \, \phi_p & -\sin \, \phi_p  \\
\sin \, \phi_p &  \cos \, \phi_p  \\
         \end{array} \right)
\left( \begin{array}{c}
\eta_N \\
\eta_S
         \end{array} \right) .
\end{eqnarray}
Here
\begin{equation}\label{text2}
 \bigl|\biggl. \, \eta_N \; \bigr>\biggr.\; \simeq   \;
 \frac{1}{\sqrt{2}}\;
 \bigl|\biggl. \, {\sf u\,\bar{u}} + {\sf d\,\bar{d}} \;
 \bigr>\biggr.\;, \quad
 \bigl|\biggl. \, \eta_S \; \bigr>\biggr.\; \simeq \;
 \bigl|\biggl. \, {\sf s\,\bar{s}} \; \bigr>\biggr. \,,
\end{equation}
and $\phi_p$ is the mixing angle in the nonstrange--strange quark basis.
If necessary, the model can be extended by introduction of additional
basic states, composed e.g. of a colorless pair of constituent gluons.
In the latter case one should introduce a third observable state and
proceed to $3\times3$ mixing matrix.

\begin{figure}[t]
\hbox{ \hspace*{50pt}
       \epsfxsize=0.75\textwidth \epsfbox{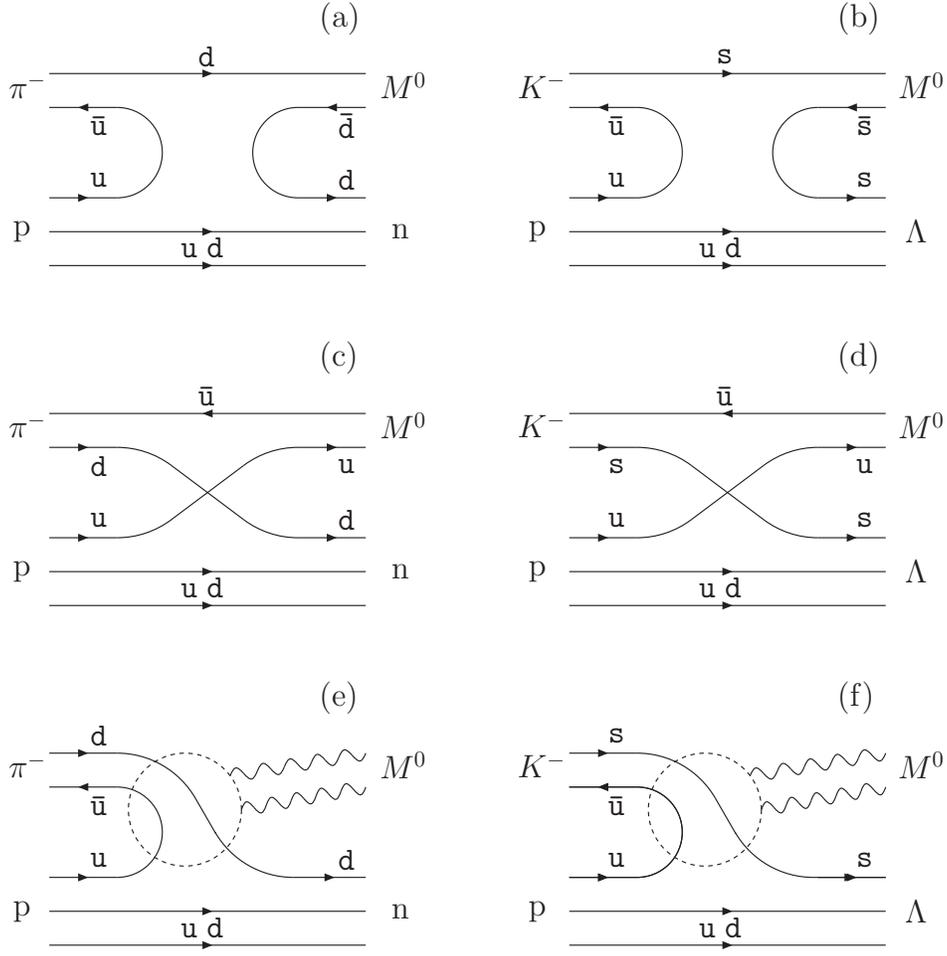}}
\caption{ Diagrams for processes $\pi^{-}\,p \to M^0\,n$ (a,c,e) and
$K^{-}\,p \to M^0\Lambda$ (b,d,f) in the na\"{\i}ve quark model. $M^0$
means any isosinglet pseudoscalar state. Dotted circles symbolically
indicate areas of formation of constituent gluons.}
\label{Figur1}
\end{figure}

In the framework of this model the production of $\eta$ and $\eta'$, or
in general case of any isosinglet pseudoscalar state $M^0$, in
$\pi^{-}p$ and $K^{-}p$ charge-exchange reactions can be represented by
diagrams of Fig.~\ref{Figur1}. The upper and middle-row pairs of
diagrams in Fig.~\ref{Figur1} represent contributions with quark
annihilation and with quark exchange, respectively. The lower pair of
diagrams represents both types of contributions with simultaneous
production of colorless pair of constituent gluons. In the simplest case
the diagrams of Fig.~\ref{Figur1} exhaust the list of contributions. We
start our analysis with consideration of all mentioned diagrams, without
assumption a priori that some contributions are suppressed. In this
point we diverge from analysis \cite{GAMS-4p}, which takes into
consideration only the upper pair of diagrams of annihilation type.

First we notice that if the processes occur at high energies, then all
diagrams of Fig.~\ref{Figur1} imply hard subprocesses. Really, in all
cases the quark or antiquark, being a part of high-energy $\pi^{-}$ or
$K^{-}$, interacts with a quark of proton in the fixed target. As a
result either a large energy is released, resulting in the production of
quark-antiquark pair with one of the quarks flying away with high energy
and another remaining in the target, or the incident quark join the
target knocking out another quark with high energy, see
Fig.~\ref{Figur1}(a,b) and Fig.~\ref{Figur1}(c,d), respectively. In
diagrams Fig.~\ref{Figur1}(e,f) both processes occur with the difference
that the energy is mainly transferred to the pair of constituent gluons
rather than to a quark or antiquark. So, since all subprocesses include
hard component, they can be considered from the standpoint of
perturbative QCD. Actually all they arise in the leading order in the
coupling constant $\alpha_s$. In the upper two pairs of diagrams this is
obtained by joining two internal quark lines via a gluon line (see
Fig.~\ref{Figur2} below). In the case of the lower pair the
corresponding diagrams are obtained by joining of one gluon to the line
with quark annihilation and of another gluon to the line with quark
exchange.

In this manner all subprocesses in Fig.~\ref{Figur1} arise in the common
order in $\alpha_s$. It can easily be seen that they arise in the common
order in the $N_{c}^{-1}\,$-expansion, as well, where $N_{c}$ is the
number of colors. For the upper two pairs of diagrams this is obvious
in view of the fact that they are planar.\footnote{The diagrams of the
middle-row pair are reduced to explicitly planar form by flipping
over on $180^0$ of the upper parts of diagrams. This gives topologically
equivalent representation of the same diagrams~\cite{tHooft}.} In the
case of the lower pair of diagrams this can be seen taking into
consideration the boundary condition under which the diagrams must be
considered. Specifically, one must equate the colors of the initial and
final quarks in the cases of mesons and barions (in the lower pair of
diagrams, of barions only) and consider colorless combinations of quarks
(gluons) in mesonic states.

So, both in terms of the counting in powers of $\alpha_s$ and the
$N_{c}^{-1}\,$-expansion all diagrams of Fig.~\ref{Figur1} should have
comparable contributions. Nevertheless, the contributions of the lower
pair of diagrams are strongly suppressed. This follows from the fact
that the lower diagrams correspond to double parton processes, and
contributions of such processes amount to about 5\% of contributions of
single parton processes \cite{DPP1,DPP2,DPP3}. 

Thereby we are coming to consideration of only two pairs of diagrams in
the cases of both reaction. Under the assumption that they occur at high
energies, for their description the approach of the parton model is
applicable. In relation to the GAMS-4p data \cite{GAMS-4p} we consider
this is justified as the data were collected at $\sqrt{s} \approx 8$
GeV. Taking into account the actual distributions of valence quarks in
mesons and in proton \cite{Meson,Proton} (at close energy scale) this in
average corresponds to $\sqrt{\hat{s}} \approx 2.2$ GeV for the hard
subprocesses, which is sufficient for application of the parton model.

Further in this analysis we will be interested in phenomenon of mixing
of states only. This prescribes us to consider the processes in the
limit of zero transfers. Thereby we escape contributions due to
exchanges in the $t$-channel and thus get opportunity for meaningful
description of relative quark content in final states. Notice, at high
energies additionally the mass differences and kinematic effects are
irrelevant. Earlier the mixing phenomenon on the basis of
charge-exchange reactions was usually studied in the limit of zero
transfers, as well \cite{GAMS-4p,NICE}.

Concluding this section we note that subprocesses due to exchanges by
soft gluons, in particular in the $t$-channel, give significant
contributions, too. Actually they are even stronger than the hard
subprocesses, and it is impossible to describe them by perturbative
methods. However, they do not affect the formation of flavor in the
final state. We will take into account such contribution in the spirit
of factorization hypothesis, i.e.~we will assume that the soft
contributions manifest themselves as a background on which the fast
hard-scaterring subprocesses evolve. In other words, we will describe
the cross-section as the product of the hard-scattering contributions
and a soft-interaction constant. At the expense of the ``soft''
constant, we will refer also contributions arising at small momentum
fractions of the colliding particles, occurring at small $\hat{s}$ or
$\hat{u}$. Such contributions by its nature are also nonperturbative and
parton model is inapplicable for their description. Nevertheless, the
contributions referred to the ``soft'' constant must obey isotopic
invariance. At high energies one could expect also independence of the
``soft'' constant from the quark content of isosinglet mesons in the
final state. These properties are sufficient to determine the mixing of
states, if contributions of the hard subprocesses are taken into account
properly.

\begin{figure}[p]
\hbox{ \hspace*{50pt}
       \epsfxsize=0.75\textwidth \epsfbox{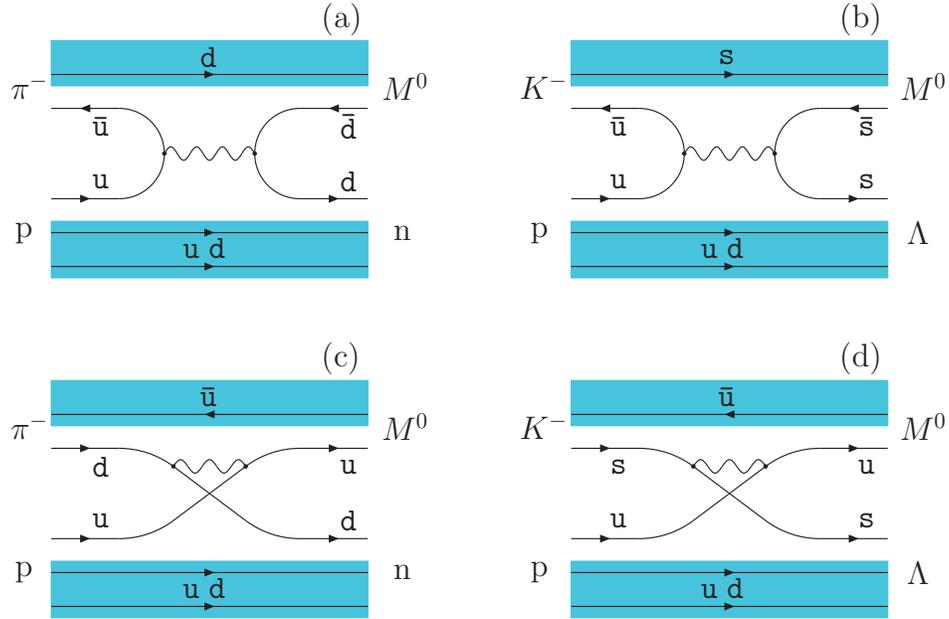}}
\caption{ Diagrams for hard subprocesses in the cases of
scattering of valence quarks. The shaded areas symbolize spectators.}
\label{Figur2}
\end{figure}
\begin{figure}[p]
\hbox{ \hspace*{50pt}
       \epsfxsize=0.75\textwidth \epsfbox{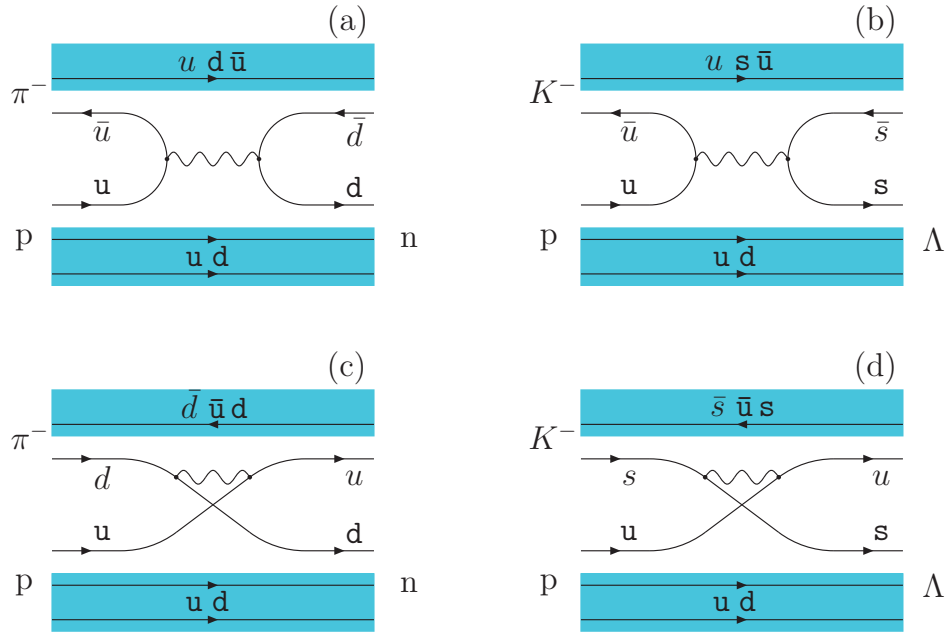}}
\caption{ Diagrams for the hard subprocesses in the cases of
scattering of sea quarks. The shaded areas symbolize spectators. The sea
quarks are indicated by slanted letters. }
\label{Figur3}
\end{figure}

\section{Analysis of partonic contributions}\label{sec3}

Now we proceed to the analysis of the hard-scattering subprocesses. In
the cases of scattering of the valence and sea quarks the corresponding
diagrams are presented by Fig.~\ref{Figur2} and Fig.~\ref{Figur3},
respectively. At first we note that at zero transfers the quarks in the
final state are distributed in momenta exactly as the relevant quarks
in the initial state are distributed. Really, since partons are
massless, in the c.m.s.~of colliding partons their momenta after the
collision are equal to the momenta before the collision. Clearly, the
same is true in the c.m.s.~of real particles. Then, repeating the
thought ``experiment'' for measuring the momenta of scattered partons,
we conclude that the relevant partons in the initial and final states
have identical distributions. So if a valence quark is scattered, then
the appropriate quark in the final state can be considered as the
valence quark, again. Correspondingly, a sea quark turns into a sea
quark. The parton distributions in spectator are unchanged, too. In the
case of valence quark scattering this means that the resulting beam of
partons is equivalent to that of a real meson. Therefore this parton
beam should transform into an observable state of isosinglet meson. The
probability of this event is determined by the proportion in which in
this meson a given set of valence quarks is presented.

In the case of scattering of sea quark situation is drastically
different. Really, as follows from Fig.~\ref{Figur3}, the sea quark in
the final state appears with a flavor that is not compensated by flavors
of other sea quarks. So the partons in the final state are incorrectly
distributed, not as required in one-particle state. This means that in
the result of hard-scattering some intermediate excitation arises. In
the course of the final hadronization it must decay to multiparticle
state. This means that the scattering of sea quarks do not contribute to
the reactions under consideration. Thus we come to the conclusion that
in the limit of zero transfers the charge-exch\-ange reactions can be
described in the valence approximation, in which the presence of sea
quarks is ignored. Further we exclude subprocesses with sea quarks from
consideration.

The above observation allows us to estimate relative contributions of
the hard-scattering subprocesses. For this purpose we calculate
amplitudes squared for subprocesses of Fig.~\ref{Figur2}. (Notice, they
do not interfere with each other.) In doing so we keep in mind
that the presence of spectators imply certain conditions at
calculating the amplitudes. Namely, the color indices of quarks must
coincide pairwise in the cases of mesons and baryons, and in the case of
mesons the helicity of the quark in the final state must coincide with
the helicity of corresponding quark in the initial state, in order to
compensate helicity of the spectator. Given these conditions, direct
calculations lead to the following results:
\begin{equation}\label{text3}
 |M_{QA}|^2 = N \times \hat{u}^2/\hat{s}^2\,,
 \qquad
 |M_{QE}|^2 = N \times \hat{s}^2/\hat{u}^2 \,.
\end{equation}
Here $M_{QA}$ and $M_{QE}$ are the amplitudes for diagrams with quark
annihilation and with quark exchange, respectively, $\hat{s}$ and
$\hat{u}$ are Mandelstam variables for subprocesses, $N$ in both cases
is the one and the same constant proportional to $\alpha_s^2\,$. At
$\hat{t}=0$ we have $\hat{u}=-\hat{s}$. So in the limit of zero
transfer the amplitudes squared in (\ref{text3}) coincide each other.
Consequently the contributions of the hard-scattering subprocesses to
the cross-section of the whole of process coincide, as well.

\section{The $\eta - \eta'$ mixing}\label{sec4}

The above result is the key to further analysis. Based on it and in view
of Section \ref{sec2}, we immediately conclude that at high energies and
zero transfers the following relations take place:
\begin{equation}\label{text4}
\sigma(\pi^- {\rm p} \to \eta_N \, {\rm n}) \;=\; \sigma_{\pi}\;, \qquad
\sigma(\pi^- {\rm p} \to \eta_S \, {\rm n}) \;=\; 0\,,
\end{equation}
\begin{equation}\label{text5}
\sigma(K^- {\rm p} \to \eta_S \, \Lambda) \;=\;
\sigma(K^- {\rm p} \to \eta_N \, \Lambda) \;=\; \sigma_{K}\,.
\end{equation}
Hereinafter we mean ${\rm d}\sigma \!/\! {\rm d}t|_{t=0}$ under the
symbol $\sigma$.

Further, we consider the simplest scheme for $\eta-\eta'$ mixing, which
is based on assumption of completeness of two states. Namely, we assume
that the states $|\,\eta_N\!> $ and $|\,\eta_S\!>$ in (\ref{text2}) can
be considered as superpositions of only two observables states of $\eta$
and $\eta'$. From (\ref{text4}) with taking into account (\ref{text1})
and (\ref{text2}), we get
\begin{equation}\label{text6}
 \sigma(\pi^- {\rm p} \to \eta'   {\rm n}) \;=\;
 \sigma_{\pi}  \, \sin^2\phi_P \,,
\end{equation}
\begin{equation}\label{text7}
 \sigma(\pi^- {\rm p} \to \,\eta \, {\rm n}) \;=\;
 \sigma_{\pi}  \, \cos^2\phi_P \,.
\end{equation}
The ratio of (\ref{text6}) and (\ref{text7}) is
\begin{equation}\label{text8}
 R_{\pi}^{\eta'/\eta} \;=\; \tan^2 \phi_P\,.
\end{equation}
Analogically, with taking into account factor $1/\!\sqrt{2}$ in
(\ref{text2}), from relations (\ref{text5}) we get
\begin{equation}\label{text9}
\textstyle \sigma(K^- {\rm p} \to \eta' \Lambda) \;=\; \sigma_{K}  \,
\left( \frac{1}{2}\sin^2\phi_P \;+\; \cos^2\phi_P \right)\,,
\end{equation}
\begin{equation}\label{text10}
\textstyle \sigma(K^- {\rm p} \to \eta \, {\rm n}) \;=\; \sigma_{K} \,
\left( \frac{1}{2}\cos^2\phi_P \;+\; \sin^2\phi_P \right)\,.
\end{equation}
The corresponding ratio of the cross-sections is
\begin{equation}\label{text11}
 R_{K}^{\eta'/\eta} \;=\;
\frac{\frac{1}{2}\tan^2\phi_p + 1}{\frac{1}{2} + \tan^2\phi_p}\,.
\end{equation}
It is worth mentioning that relation (\ref{text11}) is also valid in the
cases of reactions $K^- {\rm p} \to \eta'(\eta) \Sigma^0$ and $K^- {\rm
n} \to \eta'(\eta) \Sigma^-$.

Now we proceed to the application of our results to the GAMS-4$\pi$
data. We use the following values for the differential cross-section
ratios at $t=0$ \cite{GAMS-4p}:
\begin{eqnarray}\label{text12}
 R_{\pi}^{\eta'/\eta} &=& 0.54 \pm 0.04\,,
\\[0.5\baselineskip]\label{text13}
 R_{K}^{\eta'/\eta}   &=& 1.27 \pm 0.15\,.
\end{eqnarray}
On the basis of (\ref{text8}) and (\ref{text12}), we derive
\begin{equation}\label{text14}
 \phi_p = (36.3 \pm 1.0)^0\,.
\end{equation}
On the basis of (\ref{text9}) and (\ref{text13}), we get
\begin{equation}\label{text15}
 \phi_p = (34.6 \pm 5.6)^0\,.
\end{equation}
In the octet-singlet representation this corresponds to
$\theta_p = -18.4^0$ and $\theta_p = -20.2^0$, respectively. Recall that
the octet-singlet basis is connected with the nonstrange-strange basis
by means of rotation on the ideal mixing angle, $\theta_p = \phi_p -
\theta_i$, $\theta_i = \arctan\!\sqrt{2}$ ($\theta_i \approx 54.7^0$).

Comparing (\ref{text14}) with (\ref{text15}), we see that the values for
the mixing angle obtained on the basis of independent data for $\pi^-$
and $K^-$ beams, excellently coincide with each other. This means that
the description of the mixing of $\eta$ and $\eta'$ does not require an
introduction of any states that do not fit into the scheme of basic
states $|\eta_N\!\!>$ and $|\eta_S\!\!>$. The average value of the
mixing angle, we determine by the average value of the tangent. From
(\ref{text12}), (\ref{text13}), (\ref{text8}), (\ref{text11}), we get
$\tan \phi_p =0.73 \pm 0.03$. So the resultant angle is
\begin{equation}\label{text16}
\phi_p = (36.1 \pm 0.9)^0\,.
\end{equation}
This corresponds to $\theta_P =  (-18.7 \pm 0.9)^0$ in the
octet-singlet representation. The latter result can be compared with
that of NICE experiment $\theta_p= (-18.2 \pm 1.4)^0$ \cite{NICE}.

\section{Conclusion}

In this paper an approach is proposed to describe processes of
charge-exchanges of $\pi^{-}$ and $K^{-}$ on protons at high energies
in the limit of zero transfers, based on the conceptions of the parton
model. Specificity of application of the parton model is that partons
after the hard scattering do not leave independently the interaction
region, but join the spectator. In fact, this behavior of partons is
determined by kinematic conditions under which the processes are
considered. However, on the other hand, effectively this implies some
restrictions imposed on hard subprocesses. Namely, the scattering of
only valence quark leads to production of observable one-particle
states, while scattering of sea quark leads to disruption of the final
state and ultimately formation of multiparticle state. In essence this
means validity of the valence approximation for the description of
processes under consideration. On the basis of this
property, we have shown equality of contributions with annihilation and
exchange of quarks in the hard-scattering subprocesses. This fact was
omitted in the previous analysis \cite{GAMS-4p}, and contributions of
the latter type were not taken into consideration. However such
contributions are important in the case of scattering of $K^-$, as they
open an additional channel for production of $\eta$ and $\eta'$ in the
final state. Finally, on the base of GAMS-4$\pi$ data \cite{GAMS-4p}, we
have shown that the description of mixing of $\eta$ and $\eta'$ does not
require an introduction of any states that do not fit into the scheme of
basis states $|\eta_N\!\!>$ and $|\eta_S\!\!>$. The obtained value of
the mixing angle is $\phi_p=(36.1\pm 0.9)^0$.

\end{document}